\pdfoutput=1

\documentclass[11pt]{article}

\usepackage[final]{acl}

\usepackage{times}
\usepackage{latexsym}

\usepackage[T1]{fontenc}
\usepackage[utf8]{inputenc}

\usepackage{microtype}

\usepackage{inconsolata}

\usepackage{algorithm}
\usepackage{algorithmic}
\usepackage{multirow}
\usepackage{booktabs}
\usepackage{amsmath}
\usepackage{amssymb}
\usepackage{graphicx}

\usepackage{newfloat}
\usepackage{listings}

\title{TCSinger: Zero-Shot Singing Voice Synthesis with Style Transfer\\ and Multi-Level Style Control}

\author{%
\normalsize
Yu Zhang\quad
Ziyue Jiang\quad
Ruiqi Li\quad
Changhao Pan\quad
Jinzheng He\quad
\\\normalsize
\textbf{Rongjie Huang}\quad
\textbf{Chuxin Wang}\quad
\textbf{Zhou Zhao}\thanks{Corresponding Author}\quad
\\
Zhejiang University, Shanghai AI Laboratory\\
\texttt{\{yuzhang34,ziyuejiang,zhaozhou\}@zju.edu.cn}
}

\begin{document}
\maketitle
\begin{abstract}

Zero-shot singing voice synthesis (SVS) with style transfer and style control aims to generate high-quality singing voices with unseen timbres and styles (including singing method, emotion, rhythm, technique, and pronunciation) from audio and text prompts. 
However, the multifaceted nature of singing styles poses a significant challenge for effective modeling, transfer, and control. 
Furthermore, current SVS models often fail to generate singing voices rich in stylistic nuances for unseen singers. 
To address these challenges, we introduce TCSinger, the first zero-shot SVS model for style transfer across cross-lingual speech and singing styles, along with multi-level style control.
Specifically, TCSinger proposes three primary modules:
1) the clustering style encoder employs a clustering vector quantization model to stably condense style information into a compact latent space;
2) the Style and Duration Language Model (S\&D-LM) concurrently predicts style information and phoneme duration, which benefits both;
3) the style adaptive decoder uses a novel mel-style adaptive normalization method to generate singing voices with enhanced details.
Experimental results show that TCSinger outperforms all baseline models in synthesis quality, singer similarity, and style controllability across various tasks, including zero-shot style transfer, multi-level style control, cross-lingual style transfer, and speech-to-singing style transfer.
Singing voice samples can be accessed at \url{https://aaronz345.github.io/TCSingerDemo/}.
Code can be found at \url{https://github.com/AaronZ345/TCSinger}.

\end{abstract}

\section{Introduction}

Singing Voice Synthesis (SVS) aims to generate high-quality singing voices using lyrics and musical notations, attracting broad interest from both industry and academic communities.
The pipeline of traditional SVS systems involves an acoustic model to transform musical notations and lyrics into mel-spectrograms, which are subsequently synthesized into the target singing voice using a vocoder.

Recent years have seen significant advancements in SVS technology \citep{shi2022muskits, cho2022mandarin, zhang2023wesinger, kim2024adversarial, liu2022diffsinger,zhang2024stylesinger}. 
However, the growing demand for personalized and controllable singing experiences presents challenges for current SVS models. 
Unlike traditional SVS tasks, zero-shot SVS with style transfer and style control seeks to generate high-quality singing voices with unseen timbres and styles from audio and text prompts.
This approach can be extended to more personalized and controllable applications, such as dubbing for entertainment short videos or professional music composition.
Personal singing styles mainly include \textbf{singing method} (like bel canto), \textbf{emotion} (happy and sad), \textbf{rhythm} (including the stylistic handling of individual notes and transitions between them), \textbf{techniques} (such as falsetto), and \textbf{pronunciation} (like articulation). 
Despite this, traditional SVS methods lack the necessary mechanisms to effectively model, transfer, and control these personal styles. 
Their performance tends to decline for unseen singers, as these methods generally assume that target singers are identifiable during the training phase \citep{zhang2024stylesinger}.

Presently, zero-shot SVS with style transfer and style control primarily faces two major challenges:
1) The multifaceted nature of singing styles presents a substantial challenge for comprehensive modeling, as well as effective transfer and control. 
Previous approaches use pre-trained models to capture styles \citep{cooper2020zero}.
StyleSinger \citep{zhang2024stylesinger} uses a Residual Quantization (RQ) model to capture styles. 
However, these models focus on limited aspects of styles, neglecting styles like singing methods. 
Moreover, they fail to conduct multi-level style control.
2) Existing SVS models often fail to generate singing voices rich in stylistic nuances for unseen singers. 
VISinger 2 \citep{zhang2022visinger} uses digital signal processing techniques to enhance synthesis quality. 
Diffsinger \citep{liu2022diffsinger} employs a diffusion decoder to capture the intricacies of singing voices.
However, these methods do not adequately incorporate style information into synthesis, leading to results that lack style variations in zero-shot tasks.

To address these challenges, we introduce TCSinger, the first zero-shot SVS model for style transfer across cross-lingual speech and singing styles, along with multi-level style control. 
TCSinger transfers and controls styles (like singing methods, emotion, rhythm, techniques, and pronunciation) from audio and text prompts to synthesize high-quality singing voices.
To model diverse styles (like singing methods, emotion, rhythm, technique, and pronunciation), we propose the clustering style encoder, which uses a clustering vector quantization (CVQ) model to condense style information into a compact latent space, thus facilitating subsequent predictions, as well as enhance both training stability and reconstruction quality. 
For style transfer and control, we introduce the Style and Duration Language Model (S\&D-LM). The S\&D-LM incorporates a multi-task language module using audio and text prompts to concurrently predict style information and phoneme duration, thereby enhancing both. 
To generate singing voices rich in stylistic nuances, we introduce the style adaptive decoder, which employs a novel mel-style adaptive normalization method to refine mel-spectrograms with decoupled style information.
Our experimental results show that TCSinger outperforms other current best-performing baseline models in metrics including synthesis quality, singer similarity, and style controllability across various tasks, including zero-shot style transfer, multi-level style control, cross-lingual style transfer, and speech-to-singing (STS) style transfer.
Overall, our main contributions can be summarized as follows:
\begin{itemize}
\item We present TCSinger, the first zero-shot SVS model for style transfer across cross-lingual speech and singing styles, along with multi-level style control. 
TCSinger excels in personalized and controllable SVS tasks.
\item We introduce the clustering style encoder to extract styles, and the Style and Duration Language Model (S\&D-LM) to predict both style information and phoneme duration, addressing style modeling, transfer, and control.
\item We propose the style adaptive decoder to generate intricately detailed songs using a novel mel-style adaptive normalization method.
\item Experimental results show that TCSinger surpasses baseline models in synthesis quality, singer similarity, and style controllability across various tasks: zero-shot style transfer, multi-level style control, cross-lingual style transfer, and speech-to-singing style transfer.
\end{itemize}

\section{Related Works}
\subsection{Singing Voice Synthesis}

Singing Voice Synthesis (SVS) has emerged as a dynamic field focused on generating high-quality singing voices from provided lyrics and musical scores.
DiffSinger \citep{liu2022diffsinger} uses a diffusion-based decoder \citep{ho2020denoising} for high-quality generation.
VISinger 2 \citep{zhang2022visinger} uses digital signal processing techniques to enhance synthesis quality. 
\citet{kim2024adversarial} disentangles timbre and pitch using adversarial multi-task learning and improves the naturalness of generated singing voices.
\citet{choi2022melody} presents a melody-unsupervised model that only requires pairs of audio and lyrics, thus eliminating the need for temporal alignment. 
MuSE-SVS \citep{kim2023muse} introduces a multi-singer emotional singing voice synthesizer.
RMSSinger \citep{he2023rmssinger} proposes a pitch diffusion predictor to forecast F0 and UV, and a diffusion-based post-net to improve synthesis quality.
Nonetheless, these methods are based on the assumption that target singers are visible during the training phase, leading to a decline in synthesis quality in zero-shot scenarios.
For singing datasets, GTSinger \citep{zhang2024gtsinger} makes substantial contributions by releasing a multi-lingual and multi-technique annotated singing dataset. 
Recently, StyleSinger \citep{zhang2024stylesinger} has designed a normalization method to enhance the model generalization.
Furthermore, these methods do not adequately incorporate diverse style information into the synthesis of singing voices, resulting in limited style variations in generated audio for zero-shot SVS tasks.

\subsection{Style Modeling, Transfer and Control}

Modeling, transferring, and controlling styles remain pivotal areas of audio research, with past models predominantly leveraging pre-trained models to capture a limited array of styles \citep{kumar2021normalization}. 
\citet{atmaja2022evaluating} evaluates the performance of wav2vec 2.0 \citep{baevski2020wav2vec}, HuBERT \citep{hsu2021hubert}, and WavLM \citep{chen2022wavlm} in speech emotion recognition tasks. 
Generspeech \citep{huang2022generspeech} integrates global and local style adaptors to capture speech styles. 
Styler \citep{lee2021styler} separates styles into various levels of supervision.
YourTTS \citep{casanova2022yourtts} conditions the affine coupling layers of the flow-based decoder to handle zero-shot tasks. 
Mega-TTS \citep{jiang2023mega} decomposes speech into multiple attributes and models prosody using a language model. 
Recently, StyleSinger \citep{zhang2024stylesinger} has employed a residual quantization model to capture detailed styles in singing voices. 
Although these approaches have made strides in capturing some aspects of style, there remains a notable gap in fully modeling styles (like singing methods and techniques), and extending these capabilities to cross-lingual speech and singing styles, as well as in multi-level style control.

\section{TCSinger}

\begin{figure*}[t]
\centering
\includegraphics[width=1.0\textwidth]{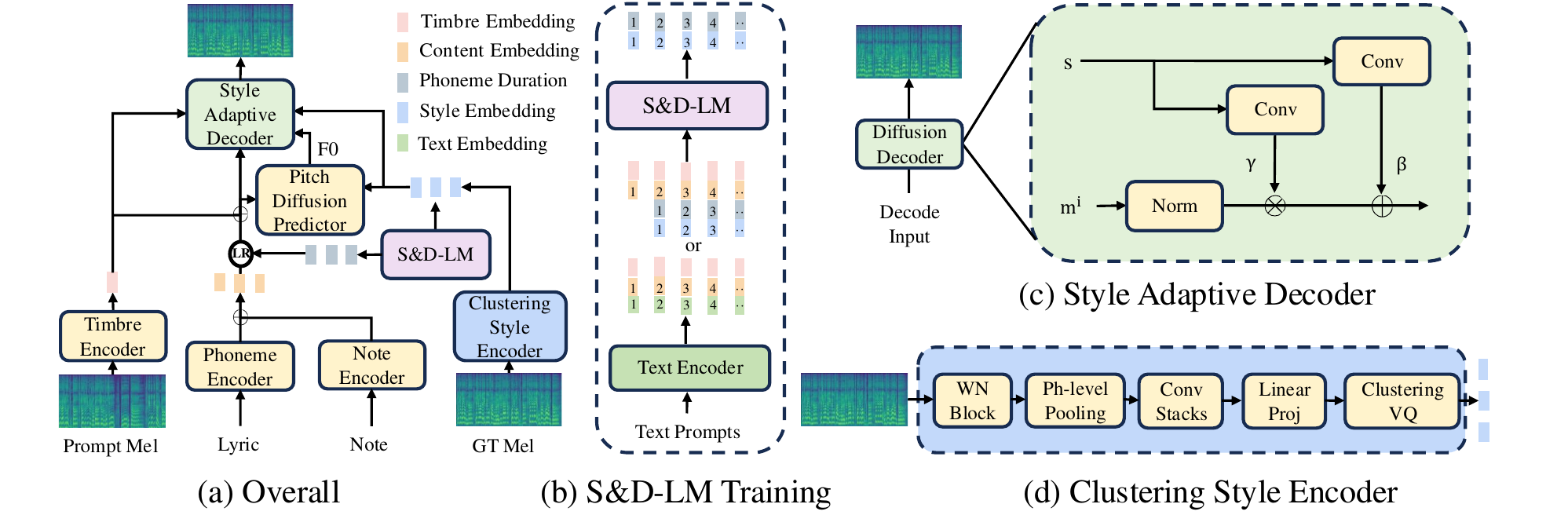}
\caption{
The architecture of TCSinger.
In Figure (a), S\&D-LM represents the Style and Duration Language Model, and LR stands for length regulator.
In Figure (b), the S\&D-LM autoregressively predicts style information and phoneme duration.
In Figure (c), intermediate mel-spectrograms are refined with style information in the style adaptive decoder.
In Figure (d), the clustering style encoder extracts style information from mel-spectrograms.
}
\label{fig: arch}
\end{figure*}

In this section, we first overview the proposed TCSinger. 
Then, we introduce several critical components, including the clustering style encoder, the style adaptive decoder, and the Style and Duration Language Model (S\&D-LM). 
Finally, we elaborate on the training and inference procedures.

\subsection{Overview}

The architecture of TCSinger is depicted in Figure \ref{fig: arch}(a).
We disentangle singing voices into separate representations for content, style (including singing method, emotion, rhythm, technique, and pronunciation), and timbre.
For content representation, lyrics are encoded through a phoneme encoder, while a note encoder captures musical notes.
For style representation, we use clustering vector quantization (CVQ) in the clustering style encoder to stably condense style information into a compact latent space, thus facilitating subsequent predictions.
For timbre representation, we feed a prompt mel-spectrogram, sampled from different audio of the same singer, into the timbre encoder to obtain a one-dimensional timbre vector, disentangling timbre from other information.
Then, we utilize the Style and Duration Language Model (S\&D-LM) to predict style information and phoneme duration.
Since styles and duration of singing voices are closely related, a composite module benefits both.
Moreover, the S\&D-LM achieves both style transfer and style control with audio and text prompts.
Next, we employ the pitch diffusion predictor for F0 prediction and the style adaptive decoder to generate the target mel-spectrogram.
The style adaptive decoder generates intricately detailed singing voices using a novel mel-style adaptive normalization method.
During training, we train the clustering style encoder for reconstruction in the first phase and S\&D-LM for style prediction in the second phase.
During inference, we can input audio prompts or text prompts to S\&D-LM for style transfer or control.
Please refer to Appendix \ref{sec: appendix1} for more details.

\subsection{Clustering Style Encoder}

To comprehensively capture styles (such as singing methods, emotion, rhythm, technique, and pronunciation) from mel-spectrograms, we introduce the clustering style encoder. 
As shown in Figure \ref{fig: arch}(d), the input mel-spectrogram is initially refined through WaveNet blocks before being condensed into phoneme-level hidden states by a pooling layer based on the phoneme boundary. 
Subsequently, the convolution stacks capture phoneme-level correlations.
Next, we use a linear projection to map the output into a low-dimensional latent variable space for code index lookup, which can significantly increase the codebook's usage \citep{yu2021vector}. 
The CVQ layer \citep{zheng2023online} then uses these inputs $x$ to generate phoneme-level style representations, establishing an information bottleneck that effectively eliminates non-style information. 
Through the dimensionality reduction of the linear projection and the bottleneck of CVQ, we achieve a decoupling of styles from timbre and content information.
Compared to traditional VQ \citep{van2017neural}, CVQ adopts a dynamic initialization strategy during training, ensuring that less-used or unused code vectors are modified more than frequently used ones, thus solving the codebook collapse issue \citep{zheng2023online}. 
To enhance training stability and improve reconstruction quality, we apply $\ell_2$ normalization to the encoded latent variables and all latent variables in the codebook.
$\ell_2$ normalization has been proven effective for VQ in the image domain \cite{yu2021vector}.
Notably, we are the first to use CVQ in the singing field, ensuring stable and high-quality extraction of style information.
We input ground truth (GT) audio during training for learning diverse styles and prompt audio during inference for style transfer.
For more details, please refer to Appendix \ref{sec: appendix1style}.

\subsection{Style Adaptive Decoder}

The dynamic nature of singing voices poses a substantial challenge to traditional mel-decoders, which often fail to effectively capture the intricacies of mel-spectrograms. 
Furthermore, using VQ to extract style information is inherently lossy \citep{razavi2019generating}, and closely related styles can easily be encoded into identical codebook indices. 
Consequently, if we employ traditional mel-decoders here, our synthesized singing voices may become rigid and lacking in stylistic variation.
To address these challenges, we introduce the style adaptive decoder, which utilizes a novel mel-style adaptive normalization method. 
While the adaptive instance normalization method has been widely used in image tasks \citep{zheng2022movq}, our work is the first to refine overall mel-spectrograms using decoupled style information. 
Our approach can infuse stylistic variations into mel-spectrograms, thereby generating more natural and diverse audio results, even when the same style quantization is used for closely related styles in decoder inputs.

As depicted in Figure \ref{fig: arch} (c), our style adaptive decoder is based on an 8-step diffusion-based decoder \citep{huang2022prodiff}. 
We utilize FFT as the denoiser and enhance it with multiple layers of our mel-style adaptive normalization method.
We denote the intermediate mel-spectrogram of the $i$-th layer in the diffusion decoder denoiser as $m^i$. 
In $i$-th layer, $m^{i-1}$ is initially normalized using a normalization method and then adapted by the scale and bias that are computed from the style embedding $s$. 
Denote the mean and standard deviation calculation as $\mu(\cdot)$ and $\sigma(\cdot)$.
We employ Layer Normalization \citep{ba2016layer} as the normalization method here.
To be more detailed, $m^i$ is given by:
\begin{equation}
\begin{aligned}
&m^i = \phi_\gamma(s)\frac{m^{i-1}-\mu(m^{i-1})}{\sigma(m^{i-1})} + \phi_\beta(s).
\end{aligned}
\label{eq: mo}
\end{equation}
$\phi_\gamma(\cdot)$ and $\phi_\beta(\cdot)$ are two learned affine transformations for converting $s$ to the scaling and bias values.
As $\phi_\gamma(\cdot)$ and $\phi_\beta(\cdot)$ inject the stylistic variant information, it encourages similar decoder inputs to generate natural and diverse mel-spectrograms.
To train the decoder, we use both the Mean Absolute Error (MAE) loss and the Structural Similarity Index (SSIM) loss \citep{wang2004image}. 
For more details, please refer to Appendix \ref{sec: appendix1dec}.

\subsection{S\&D-LM}

Singing styles (like singing methods, emotion, rhythm, technique, and pronunciation) usually exhibit both local and long-term dependencies, and they change rapidly over time with a weak correlation to content. 
This makes the conditional language model inherently ideal for predicting styles. 
Meanwhile, phoneme duration is rich in variations and closely related to singing styles.
Therefore, we propose the Style and Duration Language Model (S\&D-LM).
Through S\&D-LM, we can achieve both zero-shot style transfer and multi-level style control using audio and text prompts.

\textbf{Style Transfer}: 
Given the lyrics $\tilde{l}$, notes $\tilde{n}$ of the target, along with lyrics $l$, notes $n$, mel-spectrogram $m$ of the audio prompt, our goal is to synthesize the high-quality target singing voice's mel-spectrogram $\tilde{m}$ with unseen timbre and styles of the audio prompt. 
Initially, we use different encoders to extract the timbre information $t$, content information $c$, and style information $s$ of the audio prompt and the target content information $\tilde{c}$:
\begin{equation}
\begin{aligned}
& s=E_{style}(m),t = E_{timbre}(m), \\
& c = E_{content}(l,n), \tilde{c} = E_{content}(\tilde{l},\tilde{n}),
\end{aligned}
\label{eq: en}
\end{equation}
where $E$ denotes encoders for each attribute.
Given that the target timbre $\tilde{t}$ is anticipated to mirror the audio prompt, we also require the target styles $\tilde{s}$.
Utilizing the powerful in-context learning capabilities of language models, we design the S\&D-LM to predict $\tilde{s}$. 
Concurrently, we also use the S\&D-LM to predict the target phoneme duration $\tilde{d}$, leveraging the strong correlation between phoneme duration and styles in singing voices to enhance both predictions.
Our S\&D-LM is based on a decoder-only transformer-based architecture \citep{brown2020language}. 
We concatenate the prompt phoneme duration $d$, prompt styles $s$, prompt content $c$, target content $\tilde{c}$, and target timbre $\tilde{t}$ to form the input.
The autoregressive prediction process will be:
\begin{equation}
\begin{aligned}
&p\left(\tilde{s},\tilde{d} \mid s,d, c, \tilde{t}, \tilde{c} ; \theta \right)= \\
&\prod_{t=0}^T p\left(\tilde{s}_{t},\tilde{p}_t \mid \tilde{s}_{<t},\tilde{d}_{<t}, s,d, c, \tilde{t}, \tilde{c} ; \theta \right),
\end{aligned}
\label{eq: lm}
\end{equation}
where $\theta$ is the parameter of our S\&D-LM.

Finally, let $P$ denote the pitch diffusion predictor and $D$ for the style adaptive decoder, we can generate the target F0 and mel-spectrogram as:
\begin{equation}
\begin{aligned}
&F0=P(\tilde{s},\tilde{d},\tilde{t},\tilde{c}),\\
&\tilde{m} = D(\tilde{s},\tilde{d},\tilde{t},\tilde{c},F0).
\end{aligned}
\label{eq: de}
\end{equation}

\textbf{Style Control}: 
With alternative text prompts, we do not need to extract $s$ and $d$ from the audio prompt.
Instead, we use the text encoder to encode the global (singing method and emotion) and phoneme-level (technique for each phoneme) text prompts $tp$ to concatenate with $c$, $\tilde{c}$, and $\tilde{t}$ to form the input.
For more details about the text encoder, please refer to Appendix \ref{sec: appendix1text}.
Subsequently, the prediction process of S\&D-LM changes to:
\begin{equation}
\begin{aligned}
&p\left(\tilde{s},\tilde{d} \mid tp, c, \tilde{t}, \tilde{c} ; \theta \right)= 
\prod_{t=0}^T p\left(\tilde{s}_{t},\tilde{p}_t \mid tp, c, \tilde{t}, \tilde{c} ; \theta \right).
\end{aligned}
\label{eq: lm2}
\end{equation}

During training, we use the cross-entropy loss for style information and the Mean Squared Error (MSE) loss for phoneme duration.
For style transfer, as shown in Figure \ref{fig: arch} (b), the clustering style encoder extracts style information from the GT mel-spectrogram to train the S\&D-LM in the teacher-forcing mode.
We set a probability parameter $p$ for whether to train style transfer or style control tasks, allowing our model to handle both.

\subsection{Training and Inference Procedures}

\noindent \textbf{Training Procedures} 
The final loss terms of TCSinger in the training phase consist of the following parts: 
1) CVQ loss $\mathcal{L}_{CVQ}$: the CVQ loss for the clustering style encoder; 
2) Pitch reconstruction loss $\mathcal{L}_{gdiff}, \mathcal{L}_{mdiff}$: the Gaussian diffusion loss and the multinomial diffusion loss between the predicted and the GT pitch spectrogram for the pitch diffusion predictor; 
3) Mel reconstruction loss $\mathcal{L}_{mae}, \mathcal{L}_{ssim}$: the MAE loss and the SSIM loss between the predicted and the GT mel-spectrogram for the style adaptive decoder.
4) Duration prediction loss $\mathcal{L}_{dur}$: the MSE loss between the predicted and the GT phoneme-level duration in log scale for S\&D-LM in the teacher-forcing mode; 
5) Style prediction loss $\mathcal{L}_{style}$: the cross-entropy loss between the predicted and the GT style information for S\&D-LM in the teacher-forcing mode.

\begin{figure}[t]
\centering
\includegraphics[width=0.5\textwidth]{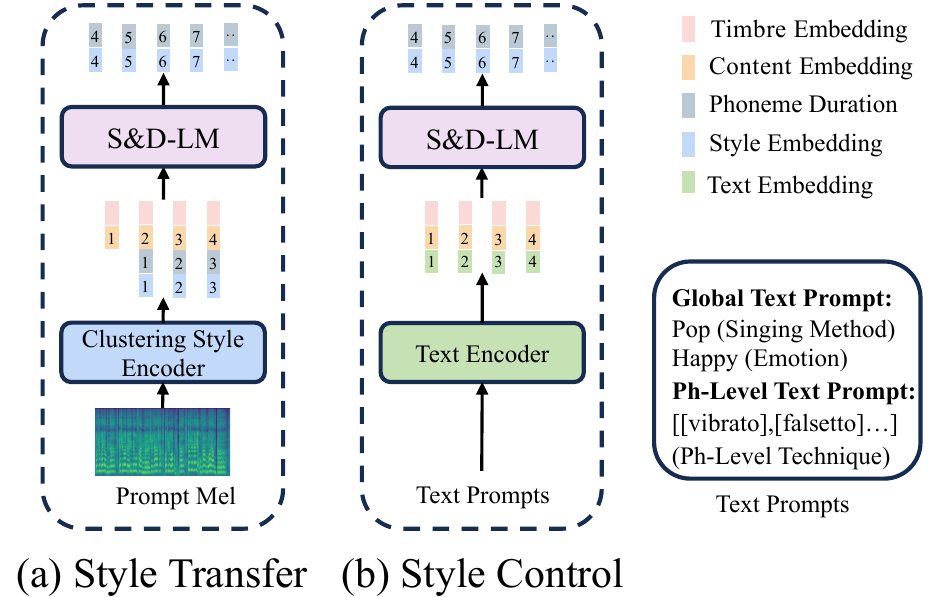}
\caption{
Inference procedure of TCSinger.
In Figure (a), the S\&D-LM extracts information from the audio prompt to predict the target style information and phoneme duration, while in Figure (b), the S\&D-LM uses multi-level text prompts to predict them. 
}
\label{fig: infer}
\end{figure}

\noindent \textbf{Inference with Style Transfer}
Refer to Figure \ref{fig: infer} (a) and Equation \ref{eq: lm}, during inference of zero-shot style transfer, we use $c$, $t$, $s$, $d$ extracted from the audio prompt, and the target content $\tilde{c}$ as inputs for the S\&D-LM, and obtain $\tilde{s},\tilde{u}$. 
Then, since the target's timbre and prompt remain unchanged, according to Equation \ref{eq: de}, we concatenate the content $\tilde{c}$, timbre $\tilde{t}$, style information $\tilde{s}$, and phoneme duration $\tilde{d}$ of the target to generate F0 by the pitch diffusion predictor, and final mel-spectrogram $\tilde{m}$ by the style adaptive decoder. 
Therefore, the generated target singing voice can effectively transfer the timbre and styles of the audio prompt. 
Moreover, we can transfer cross-lingual speech and singing styles.
For cross-lingual experiments, the language of the lyrics in the prompt and the target differ (such as English and Chinese), but the process remains the same. 
For STS experiments, speech data is used as the audio prompt, allowing the target singing voice to transfer the timbre and styles of the speech, with the remaining steps consistent. 

\noindent \textbf{Inference with Style Control}
Refer to Figure \ref{fig: infer} (b) and Equation \ref{eq: lm2}. During inference for multi-level style control, the audio prompt provides only the timbre, eliminating the need to extract prompt styles using the clustering style encoder. 
Both global and phoneme-level text prompts are encoded using the text encoder to replace $s$ and $d$, synthesizing the target $\tilde{s}$ and $\tilde{d}$, with the rest of the process consistent with style transfer tasks. 
The global text prompt encompasses singing methods (e.g., bel canto, pop) and emotions (e.g., happy, sad), while phoneme-level text prompts control techniques (e.g., mixed voice, falsetto, breathy, vibrato, glissando, pharyngeal) for each phoneme. 
Through these text prompts, we can generate singing voices with personalized timbre and independently controllable styles on both global and phoneme levels.

\section{Experiments}
\subsection{Experimental Setup}

In this section, we present the datasets utilized by TCSinger, delve into the implementation and training details, discuss the evaluation methodologies, and introduce the baseline models. 

\noindent \textbf{Dataset}
We use the open-source singing dataset with style annotations, GTSinger \citep{zhang2024gtsinger}, specifically its Chinese and English subset (5 singers, 36 hours of Chinese and English singing and speech).
We also enrich our data with M4Singer \citep{zhang2022m4singer} (20 singers, 30 hours of Chinese singing), OpenSinger \citep{huang2021multi} (93 singers, 85 hours of Chinese singing), AISHELL-3 \citep{shi2021aishell3} (218 singers, 85 hours of Chinese speech), and a subset of PopBuTFy \citep{liu2022learning} (20 singers, 18 hours of English speech and singing).
Then, we manually annotate these singing data with global style labels (singing method and emotion) and six phoneme-level singing techniques. 
Subsequently, we randomly chose 40 singers as the unseen test set to evaluate TCSinger in the zero-shot scenario for all tasks. 
Notably, our dataset partitioning carefully ensures that both training and test sets for all tasks include cross-lingual singing and speech data.
Please refer to Appendix \ref{sec: appendix2} for more details.

\noindent \textbf{Implementation Details}
We set the sample rate to 48000Hz, the window size to 1024, the hop size to 256, and the number of mel bins to 80 to derive mel-spectrograms from raw waveforms.
The default size of the codebook for CVQ is 512.
The S\&D-LM model is a decoder-only architecture with 8 Transformer layers and 512 embedding dimensions.
Please refer to Appendix \ref{sec: appendix1arch} for more details.

\noindent \textbf{Training Details}
We train our model using four NVIDIA 3090 Ti GPUs. 
The Adam optimizer is used with $\beta_1 = 0.9$ and $\beta_2 = 0.98$. 
The main SVS model takes 300k steps and the S\&D-LM takes 100k steps to train until convergence. 
Output mel-spectrograms are transformed into singing voices by a pre-trained HiFi-GAN \citep{kong2020hifi}.

\noindent \textbf{Evaluation Details}
We use both objective and subjective evaluation metrics to validate the performance of TCSinger. 
For subjective metrics, we conduct the MOS (mean opinion score) and CMOS (comparative mean
opinion score) evaluation.
We employ the MOS-Q to judge \textbf{synthesis quality} (including clarity, naturalness, and rich stylistic details), MOS-S to assess \textbf{singer similarity} (in terms of timbre and styles) between the result and prompt, and MOS-C to evaluate \textbf{style controllability} (accuracy and expressiveness of style control).
Both these metrics are rated from 1 to 5 and reported with 95\% confidence intervals. 
In the ablation study, we employ CMOS-Q to gauge synthesis quality, and CMOS-S to evaluate singer similarity.
For objective metrics, we use Singer Cosine Similarity (Cos) to judge singer similarity, and Mean Cepstral Distortion (MCD) along with F0 Frame Error (FFE) to quantify synthesis quality. 
Please refer to Appendix \ref{sec: appendix3} for more details.

\begin{table*}[t]
\centering
\small
\scalebox{1}{
\begin{tabular}{l|cc|ccc}
\toprule
\rule{0pt}{10pt} \bfseries{Method} & MOS-Q $\uparrow$ & MOS-S $\uparrow$ & FFE $\downarrow$ & MCD $\downarrow$ & Cos $\uparrow$ \\
\midrule  
GT & 4.58 $\pm$ 0.06 & / & / & / & /
\\
GT (vocoder) & 4.34 $\pm$ 0.09 & 4.39 $\pm$ 0.07 & 0.05 & 1.33 & 0.96 \\
\midrule  
YourTTS \citep{casanova2022yourtts} & 3.67 $\pm$ 0.09 & 3.76 $\pm$ 0.08 & 0.35 & 3.55 & 0.82
\\
Mega-TTS \citep{jiang2023mega} & 3.81 $\pm$ 0.08 & 3.87 $\pm$ 0.07 & 0.29 & 3.45 & 0.84
\\
RMSSinger \citep{he2023rmssinger} & 3.86 $\pm$ 0.06 & 3.8 $\pm$ 0.08 & 0.29 & 3.29 & 0.83
\\
StyleSinger \citep{zhang2024stylesinger} & 3.94 $\pm$ 0.08 & 4.01 $\pm$ 0.07 & 0.28 & 3.23 & 0.89
\\
\midrule  
TCSinger (ours) & \bf 4.12 $\pm$ 0.08 & \bf 4.28 $\pm$ 0.06 & \bf 0.22 & \bf 3.16 & \bf 0.92
\\
\bottomrule      
\end{tabular}}
\caption{
Synthesis quality and singer similarity of zero-shot style transfer. 
For subjective measurement, we employ MOS-Q and MOS-S. 
In objective evaluation, we utilize FFE, MCD, and Cos.
}
\label{tab: base}
\end{table*}

\noindent \textbf{Baseline Models}
We conduct a comprehensive comparative analysis of synthesis quality, style controllability, and singer similarity for TCSinger against several baseline models. 
Initially, we evaluate our model against the ground truth (GT) and the audio generated by HiFi-GAN (GT (vocoder)). 
Additionally, we examine TCSinger with two high-performing speech models that conduct style transfer: YourTTS \citep{casanova2022yourtts} and Mega-TTS \citep{jiang2023mega}. 
To ensure a fair comparison for singing tasks, we enhance these models with a note encoder to process music scores and train them on speech and singing data. 
Subsequently, we also compare with the best traditional SVS model, RMSSinger \citep{he2023rmssinger}.
Furthermore, we assess TCSinger's performance against StyleSinger \citep{zhang2024stylesinger}, the first model for zero-shot SVS with style transfer. 
For more details, please refer to Appendix \ref{sec: appendix3base}.

\subsection{Main Results}

\begin{figure*}[t]
\centering
\includegraphics[width=1.0\textwidth]{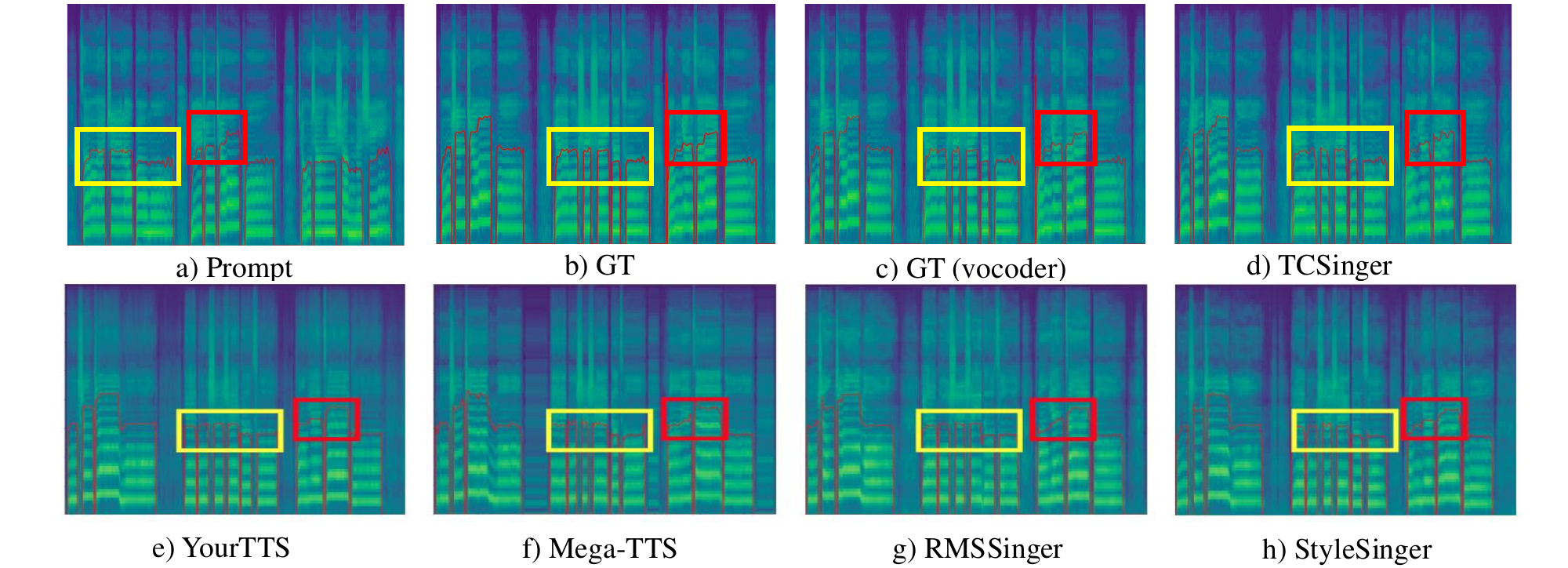}
\caption{
Mel-spectrograms depicting the results of zero-shot style transfer. 
TCSinger effectively captures the rhythm and pronunciation in red boxes, along with the vibrato technique and rhythm in yellow boxes.
}
\label{fig: svs}
\end{figure*}

\begin{table*}[t]
\centering
\small
\scalebox{1}{
\begin{tabular}{l|cccc|cc}
\toprule
\multirow{2}{*}{\bfseries{Method}} & \multicolumn{4}{c|}{\bfseries{Parallel}} & \multicolumn{2}{c}{\bfseries{Non-Parallel}}\\
& {MOS-Q $\uparrow$} & {MOS-C $\uparrow$} & {FFE $\downarrow$}  & {MCD $\downarrow$} & {MOS-Q $\uparrow$} & {MOS-C $\uparrow$} \\
\midrule
GT & 4.57$\pm$0.05 & / & / & / & / & / \\
GT (vocoder) & 4.28$\pm$0.08 & 4.31$\pm$0.09 & 0.06 & 1.35 & / & / \\
\midrule
YourTTS  & 3.59$\pm$0.11 & 3.65$\pm$0.10 & 0.38 & 3.67 & 3.55 $\pm$0.09 & 3.58$\pm$0.07 \\
Mega-TTS & 3.76$\pm$0.10 & 3.84$\pm$0.11 & 0.32 & 3.61 & 3.63$\pm$0.09 & 3.68$\pm$0.08 \\
RMSSinger & 3.83$\pm$0.06 & 3.78$\pm$0.07 & 0.31 & 3.55 & 3.69$\pm$0.03 & 3.65$\pm$0.13 \\
StyleSinger  & 3.89$\pm$0.09 & 3.93$\pm$0.11 & 0.29 & 3.45 & 3.79$\pm$0.11 & 3.85$\pm$0.10 \\
\midrule
TCSinger (ours) & \bf 4.05$\pm$0.10 & \bf 4.18$\pm$0.08 & \bf 0.24 & \bf 3.20 & \bf 3.95$\pm$0.08 & \bf 4.09$\pm$0.10 \\
\bottomrule
\end{tabular}}
\caption{
Zero-shot multi-level style control performance in both parallel and non-parallel experiments. 
For subjective measurement, we use MOS-Q and MOS-C. 
For objective measurement, we use FFE and MCD.
}
\label{tab: sc}
\end{table*}

\noindent \textbf{Zero-Shot Style Transfer}
To assess the performance of TCSinger and baseline models in the zero-shot style transfer task, we randomly select samples with unseen singers from the test set as targets and different utterances from the same singers to form prompts.
As shown in Table \ref{tab: base}, we have the following findings:
1) TCSinger exhibits outstanding synthesis quality, as indicated by the highest MOS-Q and the lowest FFE and MCD. 
This underscores the model's impressive adaptability in handling zero-shot SVS scenarios.
2) TCSinger also excels in singer similarity, as denoted by the highest MOS-S and Cos. 
This highlights our model's superior ability to model and transfer different singing styles precisely, thanks to the innovative design of our components. 
Our style adaptive decoder effectively improves the rich stylistic details of synthesis quality, rendering the singing voices more natural and of superior quality. 
Meanwhile, our clustering style encoder shows an excellent capability for modeling styles across a wide range of categories.
Finally, the S\&D-LM delivers excellent prediction results for style information and phoneme duration, significantly contributing to synthesis quality and singer similarity.
As shown in Figure \ref{fig: svs}, our TCSinger not only displays greater details in the mel-spectrogram, but also effectively learns the technique, pronunciation, and rhythm of the audio prompt. 
In contrast, other baseline models lack details in mel-spectrograms, and their pitch curves remain flat, failing to transfer diverse singing styles.
Upon listening to demos, it can be found that our model effectively transfers timbre, singing methods, emotion, rhythm, technique, and pronunciation of audio prompts.

\noindent \textbf{Multi-Level Style Control}
We add global and phoneme-level text embedding to each baseline model to enable style control. 
Then, we compare TCSinger using multi-level text prompts.
We conduct both parallel and non-parallel experiments according to the target styles.
In the parallel experiments, we randomly select unseen audio from the test set, using the GT global style and phoneme-level techniques as the target. 
In the non-parallel experiments, global styles and six techniques are randomly yet appropriately assigned.
For global styles, we specify singing methods (bel canto and pop) and emotions (happy and sad) for each test target. 
For phoneme-level styles, we select none, one or more specific techniques (mixed voice, falsetto, breathy, vibrato, glissando, and pharyngeal) for each phoneme of target content. 
As shown in Table \ref{tab: sc}, we can find that TCSinger surpasses other baseline models in both the highest synthesis quality (MOS-Q) and style controllability (MOS-C) in both parallel and non-parallel experiments. 
This indicates that, in addition to excelling in style transfer, our model also performs well in multi-level style control, and we are the first method for multi-level singing style control.
This success is attributed to our clustering style encoder's exceptional style modeling capabilities, the S\&D-LM's effective style control, and the style adaptive decoder's capacity to generate stylistically rich singing voices.
Upon listening to demos, it is obvious that our model effectively controls the global singing method and emotion, as well as phoneme-level techniques.
For more detailed results with objective evaluations, please refer to Appendix \ref{sec: appendix4sc}.

\begin{table}[t]
\centering
\small
\begin{tabular}{l|c|c}
\toprule
\bfseries{Method} & MOS-Q $\uparrow$ & MOS-S $\uparrow$ \\
\midrule
YourTTS  & 3.53 $\pm$ 0.07  & 3.59 $\pm$ 0.10 \\
Mega-TTS  & 3.71 $\pm$ 0.08 & 3.73 $\pm$ 0.09 \\
RMSSinger  & 3.75 $\pm$ 0.04  & 3.69 $\pm$ 0.09 \\
StyleSinger  & 3.85 $\pm$ 0.06 & 3.80 $\pm$ 0.07 \\
\midrule
TCSinger (ours) &\bf 3.98 $\pm$ 0.08 &\bf 4.11 $\pm$ 0.09 \\
\bottomrule   
\end{tabular}
\caption{
Synthesis quality and singer similarity comparisons for zero-shot cross-lingual style transfer. 
We use MOS-Q and MOS-S for comparison.
}
\label{tab: cross}
\end{table}

\begin{table*}[t]
\centering
\small
\begin{tabular}{l|ccccc|cc}
\toprule
\multirow{2}{*}{\bfseries{Method}} & \multicolumn{5}{c|}{\bfseries{Parallel}} & \multicolumn{2}{c}{\bfseries{Cross-Lingual}}\\
& {MOS-Q $\uparrow$} & {MOS-S $\uparrow$} & {FFE $\downarrow$}  & {MCD $\downarrow$} & {Cos $\uparrow$} & {MOS-Q $\uparrow$} & {MOS-S $\uparrow$} \\
\midrule
GT & 4.55 $\pm$ 0.06 & / & / & / & / & / & / \\
GT (vocoder) & 4.30 $\pm$ 0.07 & 4.21 $\pm$ 0.07 & 0.05 & 1.35 & 0.96 & / & / \\
\midrule
YourTTS & 3.55 $\pm$ 0.12 & 3.52 $\pm$ 0.11 & 0.39 & 3.69 & 0.80 & 3.45 $\pm$ 0.13 & 3.41 $\pm$ 0.14 \\
Mega-TTS & 3.65 $\pm$ 0.13 & 3.67 $\pm$ 0.12 & 0.36 & 3.61 & 0.82 & 3.59 $\pm$ 0.15 & 3.62 $\pm$ 0.14 \\
RMSSinger & 3.73 $\pm$ 0.07 & 3.59 $\pm$ 0.06 & 0.34 & 3.57 & 0.81 & 3.62 $\pm$ 0.04 & 3.56 $\pm$ 0.11 \\
StyleSinger & 3.80 $\pm$ 0.10 & 3.78 $\pm$ 0.11 & 0.30 & 3.46 & 0.86 & 3.68 $\pm$ 0.13 & 3.70 $\pm$ 0.12 \\
\midrule
TCSinger (ours) & \bf 3.94 $\pm$ 0.11 & \bf 4.05 $\pm$ 0.10 & \bf 0.24 & \bf 3.22 & \bf 0.90 & \bf 3.83 $\pm$ 0.12 & \bf 3.93 $\pm$ 0.11 \\
\bottomrule
\end{tabular}
\caption{
Synthesis quality and singer similarity comparisons for zero-shot speech-to-singing (STS) style transfer in both parallel and cross-lingual experiments. 
We use FFE, MCD, Cos, MOS-Q, and MOS-S for comparison.
}
\label{tab: sts}
\end{table*}

\noindent \textbf{Cross-Lingual Style Transfer}
To test the zero-shot cross-lingual style transfer performance of various models, we use unseen test data with different lyrics' languages as prompts and targets for inference (like English and Chinese), using MOS-Q and MOS-S as evaluation.
As shown in Table \ref{tab: cross}, our TCSinger outperforms other baseline models regarding both synthesis quality (MOS-Q) and singer similarity (MOS-S). 
Benefiting from our models for comprehensively modeling and effectively transferring diverse styles, TCSinger performs well in a cross-lingual environment.

\noindent \textbf{Speech-to-Singing Style Transfer}
We conducted experiments on both parallel and cross-lingual STS style transfer. 
In parallel experiments, we randomly select samples with unseen singers from the test set as targets and different speech from the same singers to form prompts.
In cross-lingual experiments, we select the speech prompt in a different lyric language from the target (such as Chinese and English).
As shown in Table \ref{tab: sts}, we can find that both synthesis quality (MOS-Q) and singer similarity (MOS-S) of TCSinger are superior to those of baseline models in both parallel and cross-lingual STS experiments. 
This demonstrates the excellent ability of our model in cross-lingual speech and singing style modeling and transfer.

\subsection{Ablation Study}

\begin{table}[t]
\centering
\small
\begin{tabular}{l|c|c}
\toprule
\bfseries{Setting} & CMOS-Q & CMOS-S \\
\midrule
TCSinger & 0.00  & 0.00 \\
\midrule
w/o CVQ & -0.25 & -0.23 \\
w/o SAD  & -0.22  & -0.18 \\
w/o DM  &  -0.12 & -0.22 \\
\bottomrule   
\end{tabular}
\caption{
Synthesis quality and singer similarity comparisons for ablation study.
SAD denotes style adaptive decoder and DM means duration model of S\&D-LM.
We use CMOS-Q and CMOS-S for comparison.
}
\label{tab: abl}
\end{table}

As depicted in Table \ref{tab: abl}, we conduct ablation studies to showcase the efficacy of various designs within TCSinger.
We use CMOS-Q to test the variation in synthesis quality, and CMOS-S to measure the changes in singer similarity.
1) Using VQ instead of CVQ in the clustering style encoder resulted in decreased synthesis quality and singer similarity, indicating the importance of CVQ for stable and high-quality style extraction.
2) Eliminating the style adaptive decoder and using an 8-step diffusion decoder \citep{huang2022prodiff} led to declines in both synthesis quality and singer similarity, underscoring the role of our method in enhancing style diversity in singing voices.
3) Predicting only styles in the S\&D-LM while using a simple duration predictor \citep{ren2020fastspeech} for phoneme duration also resulted in decreased synthesis quality and singer similarity. 
This demonstrates the mutual benefits of our model in predicting both phoneme duration and style information.
Please refer to Appendix \ref{sec: appendix4abl} for more results.

\section{Conclusion}

In this paper, we introduce TCSinger, the first zero-shot SVS model for style transfer across cross-lingual speech and singing styles, along with multi-level style control. 
TCSinger transfers and controls styles (like singing methods, emotion, rhythm, technique, and pronunciation) from audio and text prompts to synthesize high-quality singing voices.
The performance of our model is primarily enhanced through three key components: 
1) the clustering style encoder that stably condenses style information into a compact latent space using a CVQ model, thus facilitating subsequent predictions;
2) the Style and Duration Language Model (S\&D-LM), which predicts style information and phoneme duration simultaneously, which benefits both;
and 3) the style adaptive decoder that employs a novel mel-style adaptive normalization method to generate enhanced details in singing voices.
Experimental results demonstrate that TCSinger surpasses baseline models in synthesis quality, singer similarity, and style controllability across zero-shot style transfer, multi-level style control, cross-lingual style transfer, and STS style transfer. 

\section{Limitations}

Our method has two primary limitations. 
First, it currently supports control over only six singing techniques, which does not encompass the full range of commonly used singing techniques. 
Future work will focus on broadening the range of controllable techniques to enhance the versatility of style control tasks. 
Second, our multilingual data currently only facilitates cross-lingual style transfer between Chinese and English. 
In the future, we plan to gather more diverse language data for conducting multilingual style transfer experiments.

\section{Ethics Statement}

TCSinger, with its capability to transfer and control diverse styles of singing voices, could potentially be misused for dubbing in entertainment videos, raising concerns about the infringement of singers' copyrights. 
Additionally, its ability to transfer cross-lingual speech and singing styles poses risks of unfair competition and potential unemployment for professionals in related singing occupations.
To mitigate these risks, we will implement stringent restrictions on the use of our model to prevent unauthorized and unethical applications. 
We will also explore methods such as vocal watermarking to protect individual privacy.

\section*{Acknowledgements}
This work was supported by National Key R\&D Program of China (2022ZD0162000).

\bibliography{custom}

\begin{thebibliography}{39}
\expandafter\ifx\csname natexlab\endcsname\relax\def\natexlab#1{#1}\fi

\bibitem[{Atmaja and Sasou(2022)}]{atmaja2022evaluating}
Bagus~Tris Atmaja and Akira Sasou. 2022.
\newblock Evaluating self-supervised speech representations for speech emotion recognition.
\newblock \emph{IEEE Access}, 10:124396--124407.

\bibitem[{Ba et~al.(2016)Ba, Kiros, and Hinton}]{ba2016layer}
Jimmy~Lei Ba, Jamie~Ryan Kiros, and Geoffrey~E Hinton. 2016.
\newblock Layer normalization.
\newblock \emph{arXiv preprint arXiv:1607.06450}.

\bibitem[{Baevski et~al.(2020)Baevski, Zhou, Mohamed, and Auli}]{baevski2020wav2vec}
Alexei Baevski, Yuhao Zhou, Abdelrahman Mohamed, and Michael Auli. 2020.
\newblock wav2vec 2.0: A framework for self-supervised learning of speech representations.
\newblock \emph{Advances in neural information processing systems}, 33:12449--12460.

\bibitem[{Brown et~al.(2020)Brown, Mann, Ryder, Subbiah, Kaplan, Dhariwal, Neelakantan, Shyam, Sastry, Askell et~al.}]{brown2020language}
Tom Brown, Benjamin Mann, Nick Ryder, Melanie Subbiah, Jared~D Kaplan, Prafulla Dhariwal, Arvind Neelakantan, Pranav Shyam, Girish Sastry, Amanda Askell, et~al. 2020.
\newblock Language models are few-shot learners.
\newblock \emph{Advances in neural information processing systems}, 33:1877--1901.

\bibitem[{Casanova et~al.(2022)Casanova, Weber, Shulby, Junior, G{\"o}lge, and Ponti}]{casanova2022yourtts}
Edresson Casanova, Julian Weber, Christopher~D Shulby, Arnaldo~Candido Junior, Eren G{\"o}lge, and Moacir~A Ponti. 2022.
\newblock Yourtts: Towards zero-shot multi-speaker tts and zero-shot voice conversion for everyone.
\newblock In \emph{International Conference on Machine Learning}, pages 2709--2720. PMLR.

\bibitem[{Chen et~al.(2022)Chen, Wang, Chen, Wu, Liu, Chen, Li, Kanda, Yoshioka, Xiao et~al.}]{chen2022wavlm}
Sanyuan Chen, Chengyi Wang, Zhengyang Chen, Yu~Wu, Shujie Liu, Zhuo Chen, Jinyu Li, Naoyuki Kanda, Takuya Yoshioka, Xiong Xiao, et~al. 2022.
\newblock Wavlm: Large-scale self-supervised pre-training for full stack speech processing.
\newblock \emph{IEEE Journal of Selected Topics in Signal Processing}, 16(6):1505--1518.

\bibitem[{Cho et~al.(2022)Cho, Tsao, Wang, and Liu}]{cho2022mandarin}
Yin-Ping Cho, Yu~Tsao, Hsin-Min Wang, and Yi-Wen Liu. 2022.
\newblock \href {http://arxiv.org/abs/2209.10446} {Mandarin singing voice synthesis with denoising diffusion probabilistic wasserstein gan}.

\bibitem[{Choi and Nam(2022)}]{choi2022melody}
Soonbeom Choi and Juhan Nam. 2022.
\newblock A melody-unsupervision model for singing voice synthesis.
\newblock In \emph{ICASSP 2022-2022 IEEE International Conference on Acoustics, Speech and Signal Processing (ICASSP)}, pages 7242--7246. IEEE.

\bibitem[{Cooper et~al.(2020)Cooper, Lai, Yasuda, Fang, Wang, Chen, and Yamagishi}]{cooper2020zero}
Erica Cooper, Cheng-I Lai, Yusuke Yasuda, Fuming Fang, Xin Wang, Nanxin Chen, and Junichi Yamagishi. 2020.
\newblock Zero-shot multi-speaker text-to-speech with state-of-the-art neural speaker embeddings.
\newblock In \emph{ICASSP 2020-2020 IEEE International Conference on Acoustics, Speech and Signal Processing (ICASSP)}, pages 6184--6188. IEEE.

\bibitem[{He et~al.(2023)He, Liu, Ye, Huang, Cui, Liu, and Zhao}]{he2023rmssinger}
Jinzheng He, Jinglin Liu, Zhenhui Ye, Rongjie Huang, Chenye Cui, Huadai Liu, and Zhou Zhao. 2023.
\newblock Rmssinger: Realistic-music-score based singing voice synthesis.
\newblock \emph{arXiv preprint arXiv:2305.10686}.

\bibitem[{Ho et~al.(2020)Ho, Jain, and Abbeel}]{ho2020denoising}
Jonathan Ho, Ajay Jain, and Pieter Abbeel. 2020.
\newblock Denoising diffusion probabilistic models.
\newblock \emph{Advances in neural information processing systems}, 33:6840--6851.

\bibitem[{Hsu et~al.(2021)Hsu, Bolte, Tsai, Lakhotia, Salakhutdinov, and Mohamed}]{hsu2021hubert}
Wei-Ning Hsu, Benjamin Bolte, Yao-Hung~Hubert Tsai, Kushal Lakhotia, Ruslan Salakhutdinov, and Abdelrahman Mohamed. 2021.
\newblock Hubert: Self-supervised speech representation learning by masked prediction of hidden units.
\newblock \emph{IEEE/ACM Transactions on Audio, Speech, and Language Processing}, 29:3451--3460.

\bibitem[{Huang et~al.(2021)Huang, Chen, Ren, Liu, Cui, and Zhao}]{huang2021multi}
Rongjie Huang, Feiyang Chen, Yi~Ren, Jinglin Liu, Chenye Cui, and Zhou Zhao. 2021.
\newblock Multi-singer: Fast multi-singer singing voice vocoder with a large-scale corpus.
\newblock In \emph{Proceedings of the 29th ACM International Conference on Multimedia}, pages 3945--3954.

\bibitem[{Huang et~al.(2022{\natexlab{a}})Huang, Ren, Liu, Cui, and Zhao}]{huang2022generspeech}
Rongjie Huang, Yi~Ren, Jinglin Liu, Chenye Cui, and Zhou Zhao. 2022{\natexlab{a}}.
\newblock Generspeech: Towards style transfer for generalizable out-of-domain text-to-speech synthesis.
\newblock \emph{arXiv preprint arXiv:2205.07211}.

\bibitem[{Huang et~al.(2022{\natexlab{b}})Huang, Zhao, Liu, Liu, Cui, and Ren}]{huang2022prodiff}
Rongjie Huang, Zhou Zhao, Huadai Liu, Jinglin Liu, Chenye Cui, and Yi~Ren. 2022{\natexlab{b}}.
\newblock Prodiff: Progressive fast diffusion model for high-quality text-to-speech.
\newblock In \emph{Proceedings of the 30th ACM International Conference on Multimedia}, pages 2595--2605.

\bibitem[{Jiang et~al.(2023)Jiang, Ren, Ye, Liu, Zhang, Yang, Ji, Huang, Wang, Yin et~al.}]{jiang2023mega}
Ziyue Jiang, Yi~Ren, Zhenhui Ye, Jinglin Liu, Chen Zhang, Qian Yang, Shengpeng Ji, Rongjie Huang, Chunfeng Wang, Xiang Yin, et~al. 2023.
\newblock Mega-tts: Zero-shot text-to-speech at scale with intrinsic inductive bias.
\newblock \emph{arXiv preprint arXiv:2306.03509}.

\bibitem[{Kim et~al.(2023)Kim, Kim, Jun, and Kim}]{kim2023muse}
Sungjae Kim, Yewon Kim, Jewoo Jun, and Injung Kim. 2023.
\newblock Muse-svs: Multi-singer emotional singing voice synthesizer that controls emotional intensity.
\newblock \emph{IEEE/ACM Transactions on Audio, Speech, and Language Processing}.

\bibitem[{Kim et~al.(2024)Kim, Kang, and Lee}]{kim2024adversarial}
Tae-Woo Kim, Min-Su Kang, and Gyeong-Hoon Lee. 2024.
\newblock \href {http://arxiv.org/abs/2206.11558} {Adversarial multi-task learning for disentangling timbre and pitch in singing voice synthesis}.

\bibitem[{Kong et~al.(2020)Kong, Kim, and Bae}]{kong2020hifi}
Jungil Kong, Jaehyeon Kim, and Jaekyoung Bae. 2020.
\newblock Hifi-gan: Generative adversarial networks for efficient and high fidelity speech synthesis.
\newblock \emph{Advances in Neural Information Processing Systems}, 33:17022--17033.

\bibitem[{Kumar et~al.(2021)Kumar, Goel, Narang, and Lall}]{kumar2021normalization}
Neeraj Kumar, Srishti Goel, Ankur Narang, and Brejesh Lall. 2021.
\newblock Normalization driven zero-shot multi-speaker speech synthesis.
\newblock In \emph{Interspeech}, pages 1354--1358.

\bibitem[{Lee et~al.(2021)Lee, Park, and Kim}]{lee2021styler}
Keon Lee, Kyumin Park, and Daeyoung Kim. 2021.
\newblock Styler: Style factor modeling with rapidity and robustness via speech decomposition for expressive and controllable neural text to speech.
\newblock \emph{arXiv preprint arXiv:2103.09474}.

\bibitem[{Li et~al.(2024)Li, Zhang, Wang, Hong, Huang, and Zhao}]{li2024robust}
Ruiqi Li, Yu~Zhang, Yongqi Wang, Zhiqing Hong, Rongjie Huang, and Zhou Zhao. 2024.
\newblock \href {http://arxiv.org/abs/2405.09940} {Robust singing voice transcription serves synthesis}.

\bibitem[{Liu et~al.(2022{\natexlab{a}})Liu, Li, Ren, Chen, and Zhao}]{liu2022diffsinger}
Jinglin Liu, Chengxi Li, Yi~Ren, Feiyang Chen, and Zhou Zhao. 2022{\natexlab{a}}.
\newblock Diffsinger: Singing voice synthesis via shallow diffusion mechanism.
\newblock In \emph{Proceedings of the AAAI conference on artificial intelligence}, volume~36, pages 11020--11028.

\bibitem[{Liu et~al.(2022{\natexlab{b}})Liu, Li, Ren, Zhu, and Zhao}]{liu2022learning}
Jinglin Liu, Chengxi Li, Yi~Ren, Zhiying Zhu, and Zhou Zhao. 2022{\natexlab{b}}.
\newblock Learning the beauty in songs: Neural singing voice beautifier.
\newblock \emph{arXiv preprint arXiv:2202.13277}.

\bibitem[{McAuliffe et~al.(2017)McAuliffe, Socolof, Mihuc, Wagner, and Sonderegger}]{mcauliffe2017montreal}
Michael McAuliffe, Michaela Socolof, Sarah Mihuc, Michael Wagner, and Morgan Sonderegger. 2017.
\newblock Montreal forced aligner: Trainable text-speech alignment using kaldi.
\newblock In \emph{Interspeech}, volume 2017, pages 498--502.

\bibitem[{Razavi et~al.(2019)Razavi, Van~den Oord, and Vinyals}]{razavi2019generating}
Ali Razavi, Aaron Van~den Oord, and Oriol Vinyals. 2019.
\newblock Generating diverse high-fidelity images with vq-vae-2.
\newblock \emph{Advances in neural information processing systems}, 32.

\bibitem[{Ren et~al.(2020)Ren, Hu, Tan, Qin, Zhao, Zhao, and Liu}]{ren2020fastspeech}
Yi~Ren, Chenxu Hu, Xu~Tan, Tao Qin, Sheng Zhao, Zhou Zhao, and Tie-Yan Liu. 2020.
\newblock Fastspeech 2: Fast and high-quality end-to-end text to speech.
\newblock \emph{arXiv preprint arXiv:2006.04558}.

\bibitem[{Shi et~al.(2022)Shi, Guo, Qian, Huo, Hayashi, Wu, Xu, Chang, Li, Wu, Watanabe, and Jin}]{shi2022muskits}
Jiatong Shi, Shuai Guo, Tao Qian, Nan Huo, Tomoki Hayashi, Yuning Wu, Frank Xu, Xuankai Chang, Huazhe Li, Peter Wu, Shinji Watanabe, and Qin Jin. 2022.
\newblock \href {http://arxiv.org/abs/2205.04029} {Muskits: an end-to-end music processing toolkit for singing voice synthesis}.

\bibitem[{Shi et~al.(2021)Shi, Bu, Xu, Zhang, and Li}]{shi2021aishell3}
Yao Shi, Hui Bu, Xin Xu, Shaoji Zhang, and Ming Li. 2021.
\newblock \href {http://arxiv.org/abs/2010.11567} {Aishell-3: A multi-speaker mandarin tts corpus and the baselines}.

\bibitem[{Van Den~Oord et~al.(2017)Van Den~Oord, Vinyals et~al.}]{van2017neural}
Aaron Van Den~Oord, Oriol Vinyals, et~al. 2017.
\newblock Neural discrete representation learning.
\newblock \emph{Advances in neural information processing systems}, 30.

\bibitem[{Wang et~al.(2004)Wang, Bovik, Sheikh, and Simoncelli}]{wang2004image}
Zhou Wang, Alan~C Bovik, Hamid~R Sheikh, and Eero~P Simoncelli. 2004.
\newblock Image quality assessment: from error visibility to structural similarity.
\newblock \emph{IEEE transactions on image processing}, 13(4):600--612.

\bibitem[{Yu et~al.(2021)Yu, Li, Koh, Zhang, Pang, Qin, Ku, Xu, Baldridge, and Wu}]{yu2021vector}
Jiahui Yu, Xin Li, Jing~Yu Koh, Han Zhang, Ruoming Pang, James Qin, Alexander Ku, Yuanzhong Xu, Jason Baldridge, and Yonghui Wu. 2021.
\newblock Vector-quantized image modeling with improved vqgan.
\newblock \emph{arXiv preprint arXiv:2110.04627}.

\bibitem[{Zhang et~al.(2022{\natexlab{a}})Zhang, Li, Wang, Deng, Liu, Ren, He, Huang, Zhu, Chen et~al.}]{zhang2022m4singer}
Lichao Zhang, Ruiqi Li, Shoutong Wang, Liqun Deng, Jinglin Liu, Yi~Ren, Jinzheng He, Rongjie Huang, Jieming Zhu, Xiao Chen, et~al. 2022{\natexlab{a}}.
\newblock M4singer: A multi-style, multi-singer and musical score provided mandarin singing corpus.
\newblock \emph{Advances in Neural Information Processing Systems}, 35:6914--6926.

\bibitem[{Zhang et~al.(2022{\natexlab{b}})Zhang, Xue, Li, Xie, Guo, Zhang, and Gong}]{zhang2022visinger}
Yongmao Zhang, Heyang Xue, Hanzhao Li, Lei Xie, Tingwei Guo, Ruixiong Zhang, and Caixia Gong. 2022{\natexlab{b}}.
\newblock \href {http://arxiv.org/abs/2211.02903} {Visinger 2: High-fidelity end-to-end singing voice synthesis enhanced by digital signal processing synthesizer}.

\bibitem[{Zhang et~al.(2024{\natexlab{a}})Zhang, Huang, Li, He, Xia, Chen, Duan, Huai, and Zhao}]{zhang2024stylesinger}
Yu~Zhang, Rongjie Huang, Ruiqi Li, JinZheng He, Yan Xia, Feiyang Chen, Xinyu Duan, Baoxing Huai, and Zhou Zhao. 2024{\natexlab{a}}.
\newblock Stylesinger: Style transfer for out-of-domain singing voice synthesis.
\newblock In \emph{Proceedings of the AAAI Conference on Artificial Intelligence}, volume~38, pages 19597--19605.

\bibitem[{Zhang et~al.(2024{\natexlab{b}})Zhang, Pan, Guo, Li, Zhu, Wang, Xu, Lu, Hong, Wang, Zhang, He, Jiang, Chen, Yang, Zhou, Cheng, and Zhao}]{zhang2024gtsinger}
Yu~Zhang, Changhao Pan, Wenxiang Guo, Ruiqi Li, Zhiyuan Zhu, Jialei Wang, Wenhao Xu, Jingyu Lu, Zhiqing Hong, Chuxin Wang, LiChao Zhang, Jinzheng He, Ziyue Jiang, Yuxin Chen, Chen Yang, Jiecheng Zhou, Xinyu Cheng, and Zhou Zhao. 2024{\natexlab{b}}.
\newblock \href {http://arxiv.org/abs/2409.13832} {Gtsinger: A global multi-technique singing corpus with realistic music scores for all singing tasks}.

\bibitem[{Zhang et~al.(2023)Zhang, Zheng, Li, and Lu}]{zhang2023wesinger}
Zewang Zhang, Yibin Zheng, Xinhui Li, and Li~Lu. 2023.
\newblock \href {http://arxiv.org/abs/2207.01886} {Wesinger 2: Fully parallel singing voice synthesis via multi-singer conditional adversarial training}.

\bibitem[{Zheng and Vedaldi(2023)}]{zheng2023online}
Chuanxia Zheng and Andrea Vedaldi. 2023.
\newblock \href {http://arxiv.org/abs/2307.15139} {Online clustered codebook}.

\bibitem[{Zheng et~al.(2022)Zheng, Vuong, Cai, and Phung}]{zheng2022movq}
Chuanxia Zheng, Tung-Long Vuong, Jianfei Cai, and Dinh Phung. 2022.
\newblock Movq: Modulating quantized vectors for high-fidelity image generation.
\newblock \emph{Advances in Neural Information Processing Systems}, 35:23412--23425.

\end{thebibliography}

\newpage
\appendix

\section{Details of Models}
\label{sec: appendix1}

\subsection{Architecture Details}
\label{sec: appendix1arch}

We list the architecture and hyperparameters of our TCSinger in Table \ref{tab: arch}.

\begin{table}[ht]
\centering
\small
\begin{tabular*}{\hsize}{l|c|c}
\toprule
\multicolumn{2}{c|}{\bfseries{Hyper-parameter}}                         & \bfseries{Value}    \\
\midrule
\multirow{5}*{\shortstack{Phoneme\\Encoder}}            & Phoneme Embedding         & 320   \\
~                                                       & Encoder Layers            & 5     \\
~                                                       & Encoder Hidden            & 320   \\
~                                                       & Kernel Size               & 9     \\
~                                                       & Filter Size               & 1280  \\
\midrule[0.2pt]
\multirow{3}*{\shortstack{Note\\Encoder}}               & Pitches Embedding         & 320   \\
~                                                       & Type Embedding            & 320   \\
~                                                       & Duration Hidden           & 320   \\
\midrule[0.2pt]
\multirow{3}*{\shortstack{Timbre\\Encoder}}             & Encoder Layers            & 5     \\
~                                                       & Hidden Size               & 320   \\
~                                                       & Conv1D Kernel             & 31    \\
\midrule[0.2pt]
\multirow{7}*{\shortstack{Clustering\\Style\\Encoder}}  & WN Layers                 & 4     \\
~                                                       & WN Kernel                 & 3     \\ 
~                                                       & Conv Layers               & 5     \\
~                                                       & Conv Kernel               & 5     \\
~                                                       & Hidden Channel            & 320   \\
~                                                       & CVQ Embedding Size         & 512   \\
~                                                       & CVQ Embedding Channel      & 64    \\
\midrule[0.2pt]
\multirow{6}*{\shortstack{Pitch\\Diffusion\\Predictor}} & Conv Layers               & 12    \\
                                                        & Kernel Size               & 3     \\
                                                        & Residual Channel          & 192   \\
~                                                       & Hidden Channel            & 25    \\
~                                                       & Time Steps                & 100   \\
~                                                       & Max Linear $\beta$ Schedule& 0.06 \\
\midrule[0.2pt]
\multirow{4}*{\shortstack{Style\\Adapt\\Decoder}}       & Denoiser Layers           & 20    \\
~                                                       & Denoiser Hidden           & 320   \\
~                                                       & Time Steps                & 8     \\
~                                                       & Noise Schedule Type       & VPSDE \\
\midrule[0.2pt]
\multirow{6}*{\shortstack{S\&D-LM}}                     & Decoder Layers            & 8     \\
~                                                       & Style Embedding Size      & 514   \\
~                                                       & Hidden Size               & 512   \\
~                                                       & Kernel Size               & 5     \\
~                                                       & Attention Heads           & 8     \\
~                                                       & Text Embedding            & 512   \\
\midrule[0.2pt]
\multicolumn{2}{c|}{Total Number of Parameters}         & 329.5M    \\
\bottomrule
\end{tabular*}
\caption{
Hyper-parameters of TCSinger modules.
}
\label{tab: arch}
\end{table}

\subsection{Clustering Style Encoder}
\label{sec: appendix1style}

In the first phase, we train the clustering vector quantization (CVQ) codebook. 
Here, the clustering style encoder extracts style information directly from the ground truth (GT) audio. 
During the second phase of training, we train the Style and Duration Language Model (S\&D-LM) by extracting style information from the GT audio and inputting it into the S\&D-LM, facilitating training in the teacher-forcing mode. 
During style transfer inference, we use audio prompts to extract style information and then input it into the S\&D-LM.

CVQ selects encoded features as anchors to update the unused or less-used code vectors.
This strategy brings unused code vectors closer in distribution to the encoded features, increasing the likelihood of being chosen and optimized.
To train the clustering style encoder, we use the CVQ loss with $\ell_2$ normalization and the contrastive loss:
\begin{equation}
\begin{aligned}
&\mathcal{L}_{CVQ} = \|sg[\ell_2(z_e(x))] - \ell_2(e)\|^2_2 + \\
&\beta \|\ell_2(z_e(x)) - \ell_2(sg[e])\|^2_2+\mathcal{L}_{Contrastive},
\end{aligned}
\label{eq: vq}
\end{equation}
where $\text{sg}(\cdot)$ is the stop-gradient operator, $\beta$ is a commitment loss hyperparameter. 
The contrastive loss is $-\log\frac{e^{sim(e_k,\hat{z}_i^+)/\tau}}
{\sum_{i=1}^N e^{sim(e_k,\hat{z}_i^-)/\tau}}$
In particular, for each code vector $e_k$, we directly select the closest feature $\hat{z}_i^+$ as the positive pair and sample other farther features $\hat{z}_i^-$ as negative pairs using the distance computations with $\ell_2$ normalization. 
When computing the distance, we also use $\ell_2$ normalization to map all features and latent variables in the codebook onto a sphere. 
The Euclidean distance of $\ell_2$-normalized latent variables $\|\ell_2(e_k) - \ell_2(z_i)\|_2^2$ is transformed into the cosine similarity between the code vectors $e_k$ and the feature $z_i$. 
The contrastive loss effectively encourages sparsity in the codebook \citep{zheng2023online}.

\subsection{Content Encoder}
\label{sec: appendix1con}

Our content encoder is composed of a phoneme encoder and a note encoder. 
The phoneme encoder processes a sequence of phonemes through a phoneme embedding layer and four FFT blocks, culminating in the production of phoneme features. 
On the other hand, the note encoder is responsible for handling musical score information. 
It processes note pitches, note types (including rest, slur, grace, etc.), and note duration. 
Note pitches, types, and duration undergo processing through two embedding layers and a linear projection layer respectively, thereby generating note features.

\subsection{Timbre Encoder}
\label{sec: appendix1tim}

Designed to encapsulate the singer's identity, the timbre encoder extracts a global vector $t$ from the audio prompt. 
The encoder comprises several stacks of convolution layers. 
To maintain the stability of the timbre information, a one-dimensional timbre vector $t$ is obtained by averaging the output of the timbre encoder over time.

\subsection{Pitch Diffusion Predictor}
\label{sec: appendix1pit}

In our model, the pitch diffusion predictor employs a combination of both Gaussian diffusion and multinomial diffusion methodologies to generate $F0$ and $UV$ \citep{he2023rmssinger}. 
This process is described mathematically as follows:
\begin{equation}
\begin{aligned}
&q(x_t|x_{t-1}) = \mathcal{N}(x_t;\sqrt{1-\beta_t}x_{t-1},\beta_t I),\\
&q(y_t|y_{t-1}) = \mathcal{C}(y_{t}|(1-\beta_t)y_{t-1}+\beta_t/K),
\end{aligned}
\end{equation}
where $\mathcal{C}$ denotes a categorical distribution with probability parameters, $x_t \sim \{0,1\}^K$, and $\beta_t$ is the probability of uniformly resampling a category. 
In the reverse process, we train a neural network to approximate the noise $\epsilon$ from the noisy input $x_t$ and $\hat{y_0}$ from the noisy sample $y_t$ at timestep $t$. 
The equations of the reverse process are as follows:
\begin{equation}
\begin{aligned}
&E_{x_0,\epsilon}[\frac{\beta_t^2}{2\sigma_t^2\alpha_t(1-\bar \alpha_t)}||\epsilon-\epsilon_\theta(x_t,t)||],\\
&q(y_{t-1}|y_t, y_0) = \mathcal{C}(y_{t-1}|\theta_{post}(y_t, y_0)),\\ 
&\theta_{post}(y_t, y_0)=\tilde{\theta}/\sum_{k=1}^K\tilde{\theta_k},\\
&\tilde{\theta} = [\alpha_ty_t + (1-\alpha_t)/K] \odot \\
&[\bar \alpha_{t-1}y_0+(1-\bar \alpha_{t-1})/K],\\
\end{aligned}
\end{equation}
where $\alpha_t = 1-\beta_t$ and $\bar \alpha_t=\prod_{s=1}^t\alpha_s$. 
We use $p(y_{t-1}|y_t) = \mathcal{C}(y_{t-1}|\theta_{post}(y_t, \hat{y0}))$ to approximate $q(y_{t-1}|y_t, y_0)$. 
Our pitch diffusion predictor employs a non-causal WaveNet architecture for the denoiser. 
The optimization is achieved using Gaussian diffusion loss and multinomial diffusion loss.

\subsection{Style Adaptive Decoder}
\label{sec: appendix1dec}

The style adaptive decoder is based on an 8-step generator-based diffusion model \citep{huang2022prodiff}, which parameterizes the denoising model by directly predicting the clean data. 
In the training phase, we first apply Mean Absolute Error (MAE) loss.
Let $x_0$ be the original clean data, while $x_\theta$ denotes the denoised data sample:
\begin{equation}
\begin{aligned}
&\mathcal{L}_{mae} &= \left\|x_\theta\left(\alpha_t x_0 + \sqrt{1 - \alpha_t^2} \epsilon \right) - x_0\right\|,
\end{aligned}
\label{eq: mae}
\end{equation}
where $\alpha_t = \prod_{i=1}^t \sqrt{1 - \beta_i}$.
$\beta_t$ represents the predefined fixed noise schedule at diffusion step $t$. 
Additionally, $\epsilon$ is randomly sampled from a normal distribution $\mathcal{N}(0, I)$.
Furthermore, we also incorporate the Structural Similarity Index (SSIM) loss \citep{wang2004image} to the reconstruction loss:
\begin{equation}
\begin{aligned}
&\mathcal{L}_{ssim} = 1 -\\
&SSIM\left(x_\theta\left(\alpha_t x_0 + \sqrt{1-\alpha_t^2} \epsilon \right), x_0\right).
\end{aligned}
\label{eq: ssim}
\end{equation}

\subsection{Text Encoder}
\label{sec: appendix1text}

Our text encoder serves as a modular component within our framework, with a remarkably straightforward structure, akin to the type embedding model used in the note encoder. 
Our text encoder includes a global style embedding for processing global text prompts and a phoneme-level style embedding for handling phoneme-level text prompts.
Notably, the entire target should use the same singing method and emotion for naturalness, while techniques can vary between phonemes. 
Our division into global and phoneme-level styles reflects this necessity.
For global style embedding, our labeling encompasses two categories of information: two emotions (happy and sad) and two singing methods (bel canto and pop). 
We can specify these two categories, and our text encoder will process them into embedding.
For phoneme-level style embedding, each phoneme can be specified with up to six techniques. 
The techniques we used include mixed voice, falsetto, breathy, vibrato, glissando, and pharyngeal. 
We process the technique list into six technique lists with phoneme lengths and embed each separately.
Finally, we concatenate all these embeddings to form the text embedding.
During both training and inference, multi-level text prompts are thus embedded, transforming into vectors of 512 embedding size. 
This size is maintained consistent with the hidden size of the S\&D-LM, ensuring seamless integration and processing within our model architecture. 

\section{Details of Dataset}
\label{sec: appendix2}

\begin{table}[ht]
\centering
\small
\begin{tabular}{l|c|cc|cc}
\toprule
\multirow{2}{*}{\bfseries{Dataset}} & \multirow{2}{*}{\bfseries{Total/h}} & \multicolumn{2}{c|}{\textbf{Chinese}} & \multicolumn{2}{c}{\textbf{English}} \\ 
 & & \bfseries{sing} & \bfseries{speech} & \bfseries{sing}  & \bfseries{speech}\\
\midrule
GTSinger & 36 & 17 & 3 & 13 & 3 \\
M4Singer & 30 & 30 & 0 & 0 & 0 \\
OpenSinger & 85 & 85 & 0 & 0 & 0\\
AISHELL-3 & 85 & 0 & 85 & 0 & 0 \\
BuTFy & 18 & 0 & 0 & 8 & 10 \\
\midrule
\bfseries{Total/h} & 294 & 166 & 93 & 23 & 12\\
\bottomrule   
\end{tabular}
\caption{\label{tab: data}
Time distribution of our datasets for Chinese, English, speech, and singing data.
}
\end{table}

Currently, most open-source singing datasets lack music scores and multi-level style annotations. 
We use the only open-source singing dataset with style annotations GTSinger \citep{zhang2024gtsinger}, specifically its Chinese and English subset (5 singers, 36 hours of Chinese and English singing and speech). 
Additionally, we incorporate M4Singer \citep{zhang2022m4singer} (20 singers and 30 hours of Chinese singing) to expand the diversity of singers and styles. 
Subsequently, we also add OpenSinger \citep{huang2021multi} (93 singers and 85 hours of Chinese singing), AISHELL-3 \citep{shi2021aishell3} (218 singers and 85 hours of Chinese speech), and a subset of PopBuTFy \citep{liu2022learning} (20 singers, 10 hours of English speech, and 8 hours of English singing) to further expand the dataset.
None of these three datasets has music scores and alignments, so we use ROSVOT \citep{li2024robust} for coarse music score annotations and the Montreal Forced Aligner (MFA) \cite{mcauliffe2017montreal} for the coarse alignment between lyrics and audio. 
The time distribution of our datasets for cross-lingual speech and singing data are listed in Table \ref{tab: data}. 
We use all these datasets under license CC BY-NC-SA 4.0.
Moreover, with the assistance of music experts, we manually annotate part of singing data with distinct global style class labels.
We categorize songs into happy and sad based on emotion. 
In singing methods, we classify songs as bel canto and pop. 
These classifications are combined into the final style class labels, which will be the global text prompts.
We also annotate phoneme-level techniques for these singing data.
We annotate phoneme-level techniques including mixed voice, falsetto, breathy, vibrato, glissando, and pharyngeal.
These phoneme-level technique labels form the phoneme-level text prompts.
We hire all music experts and annotators with musical backgrounds at a rate of \$300 per hour. 
They have agreed to make their contributions used for research purposes.

For phonetic content, Chinese phonemes were extracted using pypinyin \footnote{https://github.com/mozillazg/python-pinyin}, English phonemes followed the ARPA standard \footnote{https://en.wikipedia.org/wiki/ARPABET}. 
We selected these standards because Chinese uses pinyin for pronunciation and ARPA includes English stress patterns, making them the most suitable phoneme standards for each language.
We then add all phonemes in a unified phoneme set.
This strategy allows our model to embed phonemes for all languages during training for all tasks.
Subsequently, we randomly chose 40 singers as the unseen test set to evaluate TCSinger in the zero-shot scenario for all tasks. 
Notably, our dataset partitioning carefully ensures that both training and test sets for all tasks include cross-lingual singing and speech data.

\section{Details of Evaluation}
\label{sec: appendix3}

\subsection{Subjective Evaluation}
\label{sec: appendix3sub}

For each task, we randomly select 20 pairs of sentences from our test set for subjective evaluation. 
Each pair consists of an audio prompt that provides timbre and styles, and a synthesized singing voice, each of which is listened to by at least 15 professional listeners.
In the context of MOS-Q and CMOS-Q evaluations, these listeners are instructed to concentrate on synthesis quality (including clarity, naturalness, and rich stylistic details), irrespective of singer similarity (in terms of timbre and styles). 
Conversely, during MOS-S and CMOS-S evaluations, the listeners are directed to assess singer similarity (singer similarity in terms of timbre and styles) to the audio prompt, disregarding any differences in content or synthesis quality (including quality, clarity, naturalness, and rich stylistic details). 
For MOS-C, the listeners are informed to evaluate style controllability (accuracy and expressiveness of style control), disregarding any differences in content, timbre, or synthesis quality (including quality, clarity, naturalness, and rich stylistic details). 
In MOS-Q, MOS-S, and MOS-C evaluations, listeners are requested to grade various singing voice samples on a Likert scale ranging from 1 to 5. 
For CMOS-Q and CMOS-S evaluations, listeners are guided to compare pairs of singing voice samples generated by different systems and express their preferences.
The preference scale is as follows: 0 for no difference, 1 for a slight difference, and 2 for a significant difference.
It is important to note that all participants are fairly compensated for their time and effort. 
We compensated participants at a rate of \$12 per hour, with a total expenditure of approximately \$300 for participant compensation.
Participants are informed that the results will be used for scientific research.

\subsection{Objective Evaluation}
\label{sec: appendix3obj}

To objectively evaluate the timbre similarity and synthesis quality of the test set, we employ three metrics: Cosine Similarity (Cos), F0 Frame Error (FFE), and Mean Cepstral Distortion (MCD). 
Cosine Similarity is used to measure the resemblance in the singer's identity between the synthesized singing voice and the audio prompt. 
This is done by computing the average cosine similarity between the embeddings extracted from the synthesized voices and the audio prompt, thus providing an objective indication of the performance in singer similarity. 
To be more specific, we use the WavLM \cite{chen2022wavlm} fine-tuned for speaker verification \footnote{https://huggingface.co/microsoft/wavlm-base-plus-sv} to extract singer embedding.
Subsequently, we use FFE, which amalgamates metrics for voicing decision error and F0 error.
FFE effectively captures essential synthesis quality information.
Next, we employ MCD for measuring audio quality:
\begin{equation}
\begin{aligned}
&\text{MCD} = \frac{10}{\ln 10} \sqrt{2 \sum_{d=1}^{D} (c_t(d) - \hat{c}_t(d))^2},
\end{aligned}
\end{equation}
where \(c_t(d)\) and \(\hat{c}_t(d)\) represent the \(d\)-th MFCC of the target and predicted frames at time \(t\), respectively, and \(D\) is the number of MFCC dimensions.

\section{Details of Baseline Models}
\label{sec: appendix3base}

In the related works section, we have described the characteristics of each baseline model and discussed their weaknesses. 
YourTTS, primarily applied to English speech, conditions the affine coupling layers of the flow-based decoder to handle zero-shot tasks. However, it does not model various styles in detail (e.g., rhythm, pronunciation) and is limited to speech, as well as lacking controllability.
Mega-TTS, which can be applied to both English and Chinese speech, decomposes speech into multiple attributes. However, it does not model various styles in detail (e.g., emotion) and is also limited to speech, lacking controllability.
RMSSinger primarily focuses on Chinese singing voices and uses a diffusion-based pitch predictor to model F0 and improve generation quality. However, it cannot perform style transfer or zero-shot SVS and lacks controllability.
StyleSinger, which primarily applies to Chinese singing, employs a residual quantization model to capture detailed styles in singing voices. However, it does not consider the styles of singing methods and techniques and also lacks controllability.

\section{Details of Results}
\label{sec: appendix4}

\subsection{Ablation Study}
\label{sec: appendix4abl}

\begin{table}[ht]
\centering
\small
\begin{tabular}{l|c|c}
\toprule
\bfseries{Metric} & \bfseries{Singer A} & \bfseries{Singer B} \\
\midrule
MOS-T  &  4.16 $\pm$ 0.08 & 2.23 $\pm$ 0.10 \\
Cos & 0.91 & 0.67 \\
\bottomrule   
\end{tabular}
\caption{
The singer similarity results from using the timbre of Singer A and the style information of Singer B to synthesize the target singing voice.
}
\label{tab: sty2}
\end{table}

To demonstrate the effectiveness of our clustering style encoder in style extraction, we conducted additional experiments. 
In these tests, we utilized the timbre of singer A and the style information of singer B. 
The results are shown in Table \ref{tab: sty2}.
Objective metrics Cosine similarity and subjective metrics MOS-T (where more than 15 professional listeners focus solely on the timbre similarity, disregarding quality and styles, ranging from 1 to 5, 5 means very similar), indicate that the synthesized results match the timbre of singer A while differing from that of singer B. 
This outcome shows that our clustering style encoder successfully decouples timbre and style in the mel-spectrogram. 

\subsection{Multi-Level Style Control}
\label{sec: appendix4sc}

Currently, there are no open-source classifiers for singing emotions or techniques to use for objective evaluation. 
Moreover, we are the first to conduct multi-level style control for singing, making the use of objective metrics quite challenging, and the accuracy of classifiers may not fully reflect the effectiveness. 
For instance, emotion in singing is relatively difficult for the model to judge, and detailed technique variations also result in low accuracy. 
Here, we provide the results of our tested emotion classifier.
We fine-tune it based on WavLM \citep{chen2022wavlm}, achieving an accuracy of 85.1\% for binary emotion classification. 
Using this, we provided the objective metric for emotion control (represented as {acc\_emo}, in \%), averaged over the test set's emotion accuracy:

\begin{table}[ht]
\centering
\small
\begin{tabular}{l|c|c}
\toprule
\bfseries{Metric} & {StyleSinger} & {TCSinger (ours)} \\
\midrule
\bfseries{acc\_emo (\%)} & 76.9\% & 79.9\% \\
\bottomrule
\end{tabular}
\caption{Emotion classification accuracy (acc\_emo) across different methods.}
\label{tab:emo_results}
\end{table}

We also test a singing method classifier based on wav2vec 2.0 \citep{baevski2020wav2vec}, achieving an accuracy of 87.3\% for the binary classification of singing methods. 
Using this, we provided the objective metric for singing method control (represented as {acc\_meth}, in \%), averaged over the test set's technique accuracy:

\begin{table}[ht]
\centering
\small
\begin{tabular}{l|c|c}
\toprule
\bfseries{Metric}& \bfseries{StyleSinger} & \bfseries{TCSinger (ours)} \\
\midrule
\bfseries{acc\_meth (\%)} & 73.1\% & 76.1\% \\
\bottomrule
\end{tabular}
\caption{Singing method classification accuracy (acc\_meth) across different methods.}
\label{tab:meth_results}
\end{table}

Technique recognition is relatively more complex. 
We design a technique recognition model based on ROSVOT \citep{li2024robust} and use cross-entropy loss for technique labels. 
The inputs of the technique recognition model include the mel-spectrogram, pitch, and phoneme boundaries, with the output being the predicted probabilities of six techniques. 
We first provide the technique recognition model's performance on our dataset:

\begin{table}[ht]
\centering
\small
\begin{tabular}{l|c|c}
\toprule
\bfseries{Tech} & \bfseries{Acc} & \bfseries{F1} \\
\midrule
mixed voice & 0.78 & 0.78 \\
falsetto & 0.84 & 0.96 \\
breathy & 0.78 & 0.99 \\
pharyngeal & 0.80 & 0.85 \\
vibrato & 0.89 & 0.70 \\
glissando & 0.85 & 0.70 \\
\bottomrule
\end{tabular}
\caption{Technique recognition model performance.}
\label{tab:tech_performance}
\end{table}

Then, we provide the objective metric for technique control:

\begin{table}[ht]
\centering
\small
\begin{tabular}{l|c|c}
\toprule
\bfseries{Metric}  & {StyleSinger} & {TCSinger (ours)} \\
\midrule
acc\_tech (\%) & 73.1\% & 76.1\% \\
\bottomrule
\end{tabular}
\caption{Technique classification accuracy (acc\_tech) for technique control.}
\label{tab:parallel_tech}
\end{table}

As shown, our TCSinger outperforms baseline models in style control tasks for any type of controllable style based on objective metrics.

\end{document}